\documentclass{aastex}
\usepackage{graphicx,epsfig}

\usepackage{emulateapj5}

\makeatletter

\newenvironment{inlinefigure}{%
\def\@captype{figure}%
\noindent\begin{minipage}{0.999\linewidth}\begin{center}}
{\end{center}\end{minipage}\smallskip}
\makeatother

\begin{document}
\title{A Redshift \lowercase{$z$} = 6.56 Galaxy Behind the Cluster Abell 370 
\altaffilmark{1}}
\author{E.~M.\ Hu,\altaffilmark{2,3} %\email{hu@ifa.hawaii.edu}
L.~L.\ Cowie,\altaffilmark{2,3} %\email{cowie@ifa.hawaii.edu}
R.~J.\ McMahon,\altaffilmark{2,4} %\email{rgm@ast.cam.ac.uk}
P.~Capak,\altaffilmark{3} %\email{capak@ifa.hawaii.edu}
F.~Iwamuro,\altaffilmark{5} %\email{iwamuro@cr.scphys.kyoto-u.ac.jp}
J.-P.~Kneib,\altaffilmark{6} %\email{kneib@ast.obs-mip.fr}
T.~Maihara,\altaffilmark{7} %\email{maihara@kuastro.kyoto-u.ac.jp}
K.~Motohara\altaffilmark{8} %\email{kmotohara@ioa.s.u-tokyo.ac.jp}
}

\altaffiltext{1}{Based in part on data collected at the Subaru Telescope,
  which is operated by the National Astronomical Society of Japan}
\altaffiltext{2}{Visiting Astronomer, W. M. Keck Observatory, which is jointly
  operated by the California Institute of Technology, the University of
  California, and the National Aeronautics and Space Administration}
\altaffiltext{3}{Institute for Astronomy, University of Hawaii,
  2680 Woodlawn Drive, Honolulu, Hawaii 96822}
\altaffiltext{4}{Institute of Astronomy, Madingley Road,
  Cambridge CB3 OHA, U.K.}
\altaffiltext{5}{Department of Physics, Kyoto University, Kitashirakawa, 
  Kyoto 606-8502, Japan}
\altaffiltext{6}{Observatoire Midi-Pyr{\'e}n{\'e}es,
14 Avenue E. Belin, 31400 Toulouse, France}
\altaffiltext{7}{Department of Astronomy, Kyoto University, Kitashirakawa, 
  Kyoto 606-8502, Japan}
\altaffiltext{8}{Institute of Astronomy, University of
  Tokyo, 2-21-1 Osawa, Mitaka, Tokyo 181-0015, Japan}

\shorttitle{A REDSHIFT $z=6.56$ GALAXY BEHIND A370}
\shortauthors{Hu et al.\/}

\slugcomment{Submitted to the Astrophysical Journal Letters}

\begin{abstract}

  We report the discovery of a redshift $z=6.56$ galaxy lying behind
  the cluster Abell 370.  The object HCM 6A was found in a
  narrowband imaging survey using a 118~\AA\ bandpass filter centered
  at 9152~\AA\ in the LRIS camera on the 10 m Keck\,II Telescope.
  Candidate Ly$\alpha$ emitters were identified by the equivalent width
  of the emission and the absence of lower wavelength flux in ultradeep
  broadband images.  HCM 6A is the first galaxy to be confirmed
  at redshift $z>6$, and has $W_{\lambda}$(observed)=190~\AA, flux =
  $2.7 \times 10^{-17}$ erg cm$^{-2}$ s$^{-1}$.  Spectra obtained with
  LRIS confirm the emission line and the continuum break across the line,
  and show an asymmetric line profile with steep fall-off on the blue side.
  Deep Subaru near-infrared CISCO images in $J$, $H$ and $K'$ which extend
  the sampled continuum to longer wavelengths give a consistent estimate of
  the continuum flux density in these line-free regions of $2.6\pm0.7 \times
  10^{-30}$ erg cm$^{-2}$ s$^{-1}$ Hz$^{-1}$.  The line width and strength,
  asymmetric profile, and very deep spectral break are only consistent
  with the interpretation of the line as a redshifted Ly$\alpha$ feature.
  From the detailed lensing model of this cluster, we estimate a lensing
  amplification of 4.5 for this galaxy, which is located slightly over
  an arcminute from the center of the cluster, for an unlensed flux of
  $6.5 \times 10^{-18}$ erg cm$^{-2}$ s$^{-1}$. The presence of such a
  galaxy suggests that the reionizing epoch is beyond $z=6.6$.

\end{abstract}


\section{Introduction}
\label{secintro}

The study of the early universe through high-redshift galaxies has seen 
substantial progress in recent times, with a number of galaxies discovered 
at redshifts $z>5$ (\citealp*{dey98}; Hu, Cowie, \& McMahon 1998; 
\citealp*{wey98}; Hu, McMahon, \& Cowie 1999; \citealp*{stern99,ellis}).
Active galaxies associated with radio galaxies \citep{wil99} have been 
seen to similar redshifts and quasars have now been detected out to $z=6.28$ 
from the Sloan survey \citep{fan00b,and01,fan01}.

The present letter describes a continuation of our narrowband filter survey 
\citep{cow98,smitty,la_lum}, which probes Ly$\alpha$ emission from 
$z=3.4\to5.7$, to a redshift $z\sim6.5$.  At $z=3$ the brightest
Ly$\alpha$ emitters have line fluxes near $10^{-16}$ erg cm$^{-2}$
s$^{-1}$ and a single, roughly 30 arcmin$^2$ field of the LRIS camera
on the Keck 10 m telescope yields a handful of objects to a flux limit
of $10^{-17}$ erg cm$^{-2}$ s$^{-1}$.  At $z\sim6.5$ the fluxes drop to
near the sensitivity limit and only occasional objects are expected to
be found \citep[e.g.,][]{thommes99}.  Therefore, to improve our
sensitivity limits we have targeted foreground massive lensing clusters
which can provide substantial amplification over a fraction of the LRIS
field. We have also observed the well studied blank fields used in
the shorter wavelength narrowband searches.  Even with the increased
sensitivity provided by the lensing detections are extremely rare and
our observations have so far yielded only one compelling detection lying
behind the cluster A370. We describe this object in the present letter.

Discriminating Ly$\alpha$ emitters from lower $z$ emission-line objects can be
difficult at lower redshifts \citep{stern99}, but as we move to higher
redshifts the continuum break signature becomes so extreme that there is
little likelihood of a misidentification provided we have sufficiently
deep continuum images.  
Songaila and Cowie \citeyearpar{gpnot} have used spectra of the
Sloan quasars to measure the average transmission of the Ly$\alpha$ forest
region as a function of redshift over the $z=4\to6$ range.  Their results
translate to a magnitude break across the Ly$\alpha$ emission line of
\begin{displaymath}
	\Delta{m} = 3.8 + 20.3\ {\rm log_{10}} \left ( {1+z}\over{7} \right )
        \label{eq:1}
\end{displaymath}
Extrapolated to $z=6.5$ this corresponds to a 4.5 magnitude break.  If the
epoch of reionization is at $z\sim 6.1$ as suggested by \citet{bec01},
the continuum blueward of Ly$\alpha$ would be essentially black.  However,
as we shall consider further in the discussion section, in this case the
radiation damping wings of the foreground neutral hydrogen extend through
the Ly$\alpha$ emission and make it unlikely we would see Ly$\alpha$
emitters at all, so the presence of the A370 emitter argues 
for a higher reionization redshift.

\begin{figure*}
\vspace*{-0.1cm}
\centerline{\epsfig{file=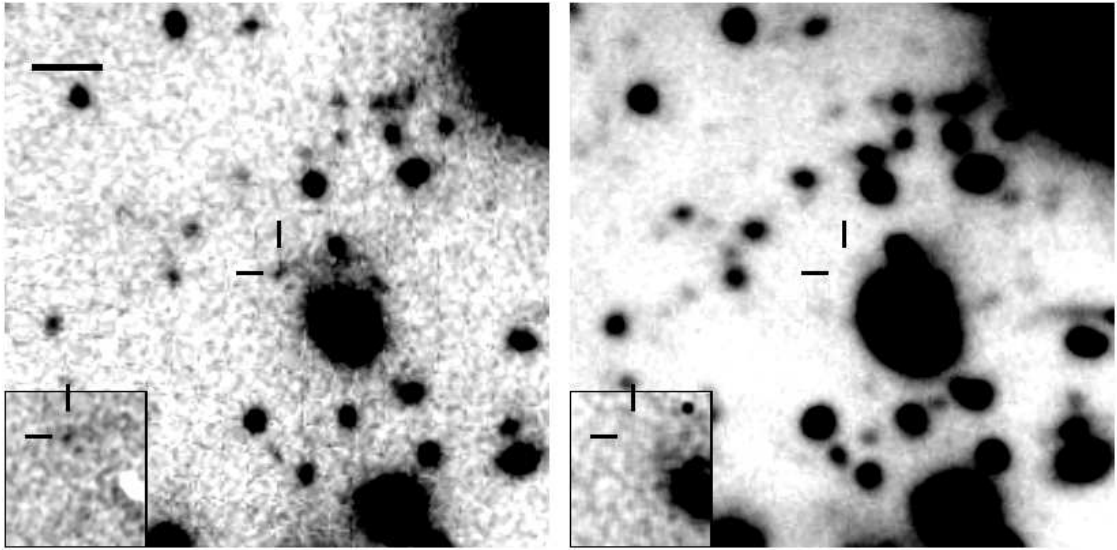,width=3.2in,angle=90}}
\caption{Narrowband 9152/118 \AA\ image (left panel) and 
   $R$-band image (right panel) of the emission-line object HCM 6A, 
   which is marked with the vertical and horizontal lines. An image
   in the narrowband filter with a normalized local continuum ($Z\/$ band)
   subtracted (inset in left panel) shows that the object, which appears
   as two fragments, is a strong emission-line object. (We have slightly
   oversubtracted the $Z\/$ band to completely remove the neighboring
   galaxy, whose core appears white.) The galaxy is not seen in the
   much deeper $R$-band image ($>8$ hours on the Keck
   10 m Telescope) or in a 5600 s F675W image taken with
   the WFPC2 camera on HST, shown in the insert in the
   right hand panel. The neighboring bright galaxy to the SW is a cluster
   member with a redshift of 0.375 \citep[``BO \#39'' in][]{mel88},
   so that the emission is not associated with this object -- in particular
   H$\alpha$ lies shortward of the filter bandpass.
   The images are centered on coordinates (J2000) 
   RA: $2^{\rm h}\,39^{\rm m}\,54\fs73$ Dec: $-1\degr\,33\arcmin\,32\farcs3$ 
   and are 37$''$ on a side. The
   bar in the upper left corner of the narrowband image shows a 5$''$ scale.
   \label{fig1:z6-compare}}
\vspace*{-0.4cm}
\end{figure*}

In Section~\ref{secdata} we summarize the narrowband observations and the 
continuum observations at longer and shorter wavelengths together with the
spectroscopic followup. The properties of HCM 6A leave little 
doubt about its identification as a high-$z$ Ly$\alpha$ emitter.  Finally, 
in the discussion we consider the implications for early star formation and
for the evolution of the intergalactic gas.

\section{Observations}
\label{secdata}

\subsection{Narrow Band Survey}
\label{secnarrow}

The present images were obtained using a 118 \AA\  filter centered on
9152 \AA\ in the LRIS camera \citep{lris} on the Keck 10 m telescopes. 
This wavelength  lies in a very dark region of the night sky  
between the OH(8-4) and OH(7-3) bands, and corresponds to Ly$\alpha$ at
a redshift of $\sim 6.5$.  In total, six fields have been imaged with this 
filter, and are summarized in Table~\ref{tbl-1}. In each case the exposures 
were made as a sequence of spatially dithered background-limited exposures 
and median skys were used to flat field the images. The images were 
calibrated with spectrophotometric standards.  The deepest exposures, 
totaling about 6 hours per field, were taken on the HDF and on fields 
centered on the clusters {A370} and \objectname{A851}. Over 
most of the area in these fields the 5 sigma limiting flux
sensitivity is approximately $1.6\times 10^{-17}$ erg cm$^{-2}$ s$^{-1}$ 
but, for the massive clusters, there
is a small area in the central regions where the lensing
amplification provides a significant gain in sensitivity.

For each field a very deep $Z\/$ band \citep[$\lambda_{\rm eff} \sim9170$
\AA;][]{la_lum} was used to measure the continuum and an extremely deep
$R$-band image was used as a shorter wavelength reference to measure
the continuum break. We searched each field for objects with observed
equivalent widths in excess of 100~\AA\ which were not visible in the $R\/$
image. Only one such object has been found in the six fields. Finding
images for HCM 6A, which lies behind the massive lensing
cluster {A370} in a region of significant amplification, are
shown in Figure~\ref{fig1:z6-compare}, which compares narrowband and deep
$R\/$ continuum images and gives coordinates.
Insets show the continuum-subtracted image and the corresponding WFPC2
F675W image from HST. (See \citet{bez99} for a description of the
wider field of the HST image).  The following subsections describe the
observations made on this object to confirm that it is a high redshift
Ly$\alpha$ emitter rather than a lower redshift emission-line object.

\subsection{Broad Band Optical Imaging}
\label{secopt}

Deep multicolor images of {A370} were obtained using LRIS
on the Keck 10 m telescopes on UT 1999 August 11, 1999 September 9--10,
2000 August 25, and 2000 December 29 and 2002 January 11--12.  The data
were taken as a sequence of dithered exposures, with net integration
times of 2400 s in $V$, 27900 s in $R$, 4050 s in $I$ and 9820 s in
$Z$. A deep $B$ (3780 s) image was obtained with ESI on Keck II on UT
2000 September 29--30.  The images were processed using median sky flats 
generated from the exposures.  Conditions were photometric during these 
observations.  The data were calibrated using the photometric and
spectrophotometric standard, HZ4 \citep{turnshek90,oke90}, and faint Landolt 
standard stars in the SA 95-42 field \citep{landolt92}.  
%
% Table 1
%
%\voffset=-0.7in
\begin{table*}
{\small
\begin{center}
\caption{Summary of LRIS Ly$\alpha$ NB9152 Survey Imaging\label{tbl-1}}
\begin{tabular}{lrrrc}
\tableline\tableline\noalign{\smallskip}
 & & & & $t_{\rm{exp}}$ \\[0.5ex]
Field & RA(2000) & Dec(2000) & {\ \ \ Exposures} & (s) \\
\noalign{\smallskip}
\tableline\noalign{\smallskip}
A370$^{\rm a}$  & 02:39:53.1 & --01:34:35 & $14\times1500$ s & 21000 \\
A851$^{\rm a}$  & 09:43:02.6 & +46:58:58 & $19\times900$ s + $7\times650$ s & 21650 \\
HDF   & 12:36:52.0 & +62:12:30 & $12\times1800$ s & 21600 \\
SSA17 & 17:06:31.1 & +43:55:40 & $6\times1200$ s & \ 7200 \\
A2390$^{\rm a}$ & 21:53:34.6 & +17:40:10 & $11\times1500$ s & 16500 \\
SSA22 & 22:17:31.9 & +00:15:01 & $9\times1500$ s & 13500 \\
\noalign{\smallskip}
\tableline
\noalign{\smallskip}
\noalign{\hrule}
\noalign{\smallskip}
\noalign{\vspace{0.05cm}}
\multispan5{~~~$^{\footnotesize\rm{a}}$ Cluster redshifts are $z=0.37$ for A370, $0.41$ for A851, and $0.231$ for A2390.\hfil}\cr
\noalign{\vspace{-0.11cm}}
\end{tabular}
\end{center}
}
\vspace*{-0.5cm}
\end{table*}
\subsection{Near-infrared Imaging}
\label{secnir}

We used the new Cooled Infrared Spectrograph and Camera for OHS 
\citep[CISCO,][]{moto98} on the Subaru 8.3\ m Telescope on UT
2000 June 18-19, July 15-16, September 10-12, and November 7 to obtain
extremely deep $J$, $H$, and $K'$ images of {A370}.
The detector used was a $1024\times 1024$ HgCdTe HAWAII array with a pixel
scale of $0.111''$.  This provides a field-of-view of $\sim 2'\times 2'$.
To minimize the image degradation, a number of sub-exposures were taken
at each position in an eight-point pentagon pattern ($5''$ step size).
Weather conditions were photometric, and the seeing was typically
$0.3''-0.5''$ during the first three observing runs, which was also
the resolution of nearly all the final reduced images. Conditions were
clear, but with variable seeing during the November observing run, with
characteristic image FWHM of $\sim0.8''$ for the A370 $H$ final image.
The data were processed using median sky flats generated from the dithered
images. The data were calibrated from observations of the UKIRT faint
standards \citep{irstds} FS27, FS29, FS6, and FS10. The total exposure
times were 13280\ s ($J$), 7680\ s ($H$), and 15360\ s ($K'$).
\begin{inlinefigure}
%\centerline{\epsfig{file=narrow_image.ps,width=3.82in,angle=0}}
\centerline{\epsfig{file=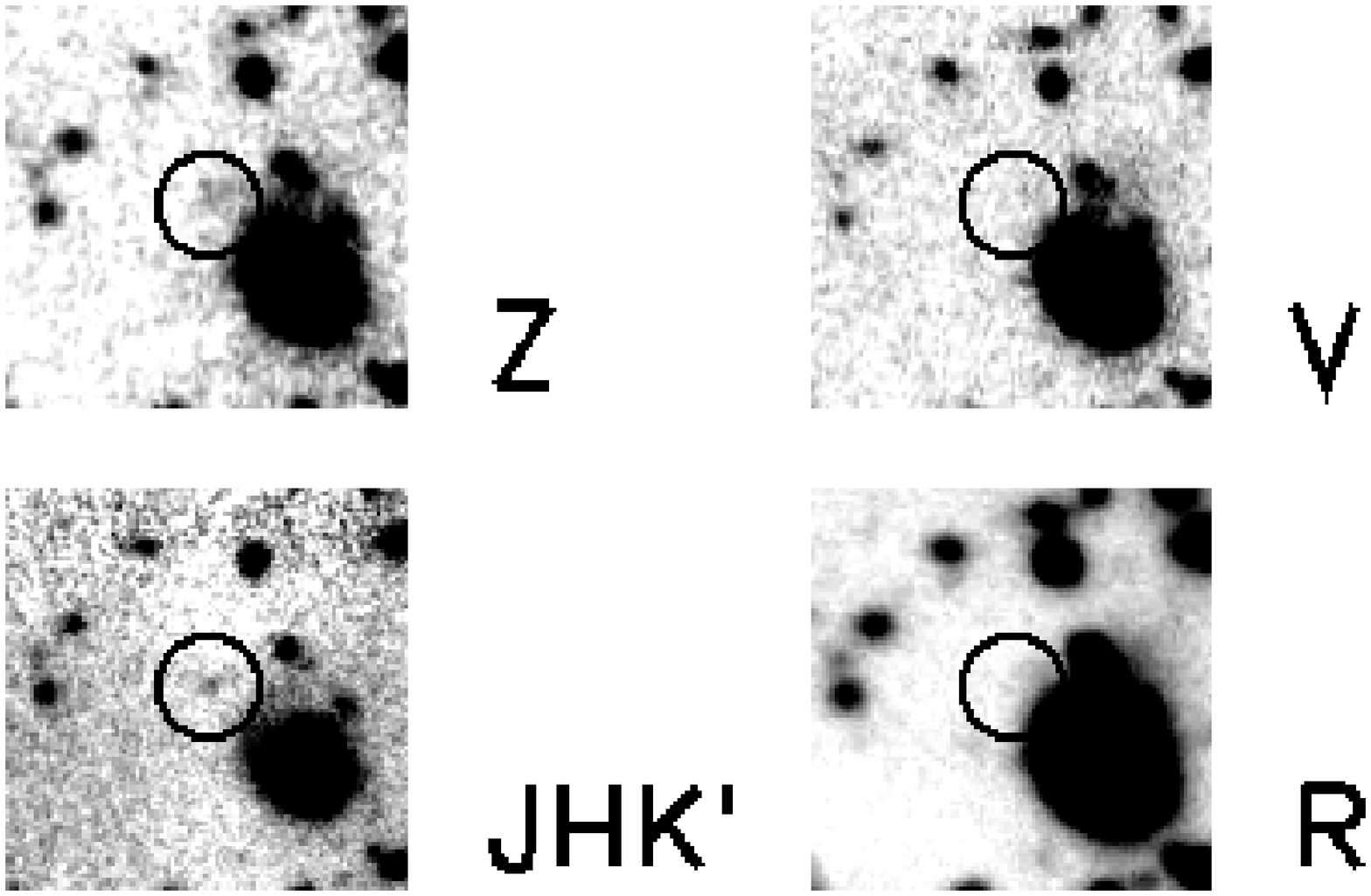,width=2.8in,angle=90,scale=0.9}}
%\centerline{\epsfig{file=z6_multi_image.eps,width=2.8in,angle=90}}
\caption{Broadband images of the redshift $z=6.56$ galaxy HCM 6A in colors
  above the Lyman break (left-hand column) and below the Lyman break (right 
  column).  The lower left image marked $JHK'$ is a weighted sum of the $J$, 
  $H$, and $K'\/$ images.  The thumbnail images are 19$''$ on a side, centered 
  on the position of the emission, and the circles are 1.2$''$ in radius.
  \label{fig2:z6-multi-image}
}
\end{inlinefigure}
\subsection{Colors and SEDs}

HCM 6A is clearly detected at all wavelengths longer than the
$Z\/$ band as is shown in Figure~\ref{fig2:z6-multi-image}, where we present
thumbnail images centered on the object, which is circled (left column).
However, it is not seen at any shorter wavelengths (right column).

The spectral energy distribution of HCM 6A is shown in
Figure~\ref{fig3:a370-6-sed}, where the error bars are 1 sigma. At wavelengths 
above the $Z\/$ band the mean flux density averaged over the $J$, $H$, and 
$K'$ bands is $2.6\pm0.7\times10^{-30}$ erg cm$^{-2}$ s$^{-1}$ Hz$^{-1}$
while at shorter wavelengths averaged over the $V\/$ and $R\/$ bands the flux 
density is $0.1\pm2.9\times10^{-31}$ erg cm$^{-2}$ s$^{-1}$ Hz$^{-1}$.  At the 
2 sigma level the continuum break exceeds 1.7 magnitudes. 

\begin{inlinefigure}
\vspace{-6pt}
\centerline{\epsfig{file=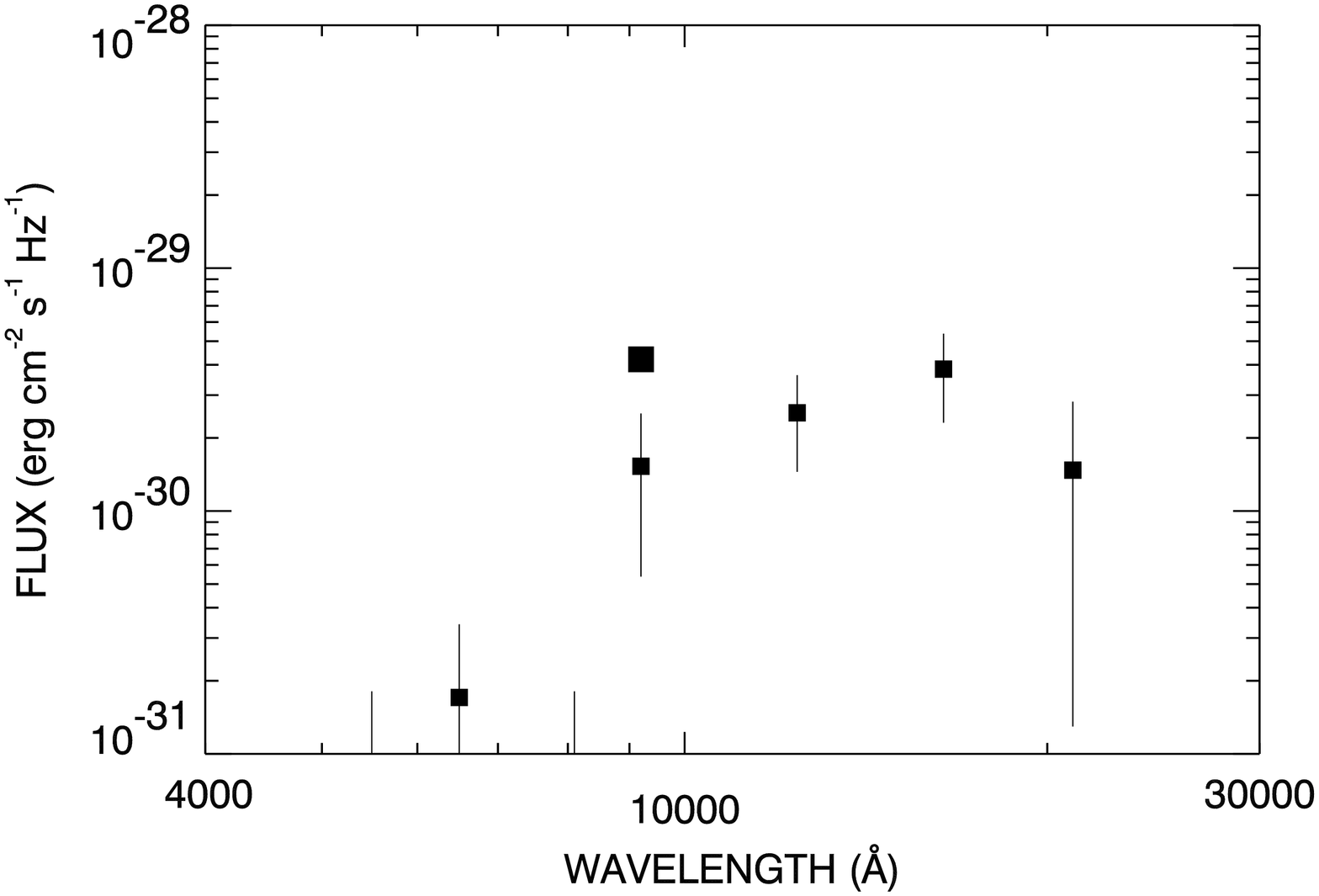,height=3.3in,angle=90}}
\vspace{-2pt}
\caption {The spectral energy distribution of HCM 6A from 4000 \AA\ to 
   22000 \AA. The broad band fluxes ($B$,$V$,$R$,$I$,$Z$,$J$,$H$,$K'$)
   are shown with small boxes while the narrowband at 9152\AA\
   is shown with the larger box. The fluxes have been measured
   in $1''$ diameter apertures centered on the object and
   the error bars are 1 sigma errors based on similar aperture
   measurements on clear portions of the field. Note the strong
   continuum break across the narrowband filter at 9152 \AA.
   \label{fig3:a370-6-sed} 
  } 
\vspace*{-0.2cm}
%\addtolength{\baselineskip}{-20pt}
\end{inlinefigure}

\subsection{LRIS Spectroscopy}

Spectra of the object were obtained with the LRIS spectrograph
during a runs in UT 16--18 September 1999, 23--24 January 2001, and
21 October 2001. The observations were made in multislit
mode using 1.4$''$ wide slits with a variety of gratings.
The emission line is clearly seen in all the spectra, and the 
stacked spectra show the line profile to be asymmetric, with
steeper blue fall-off as would be expected from
the effects of the forest absorption. The redward continuum
is most clearly seen in the spectra obtained with the 400 l/mm
grating, which in combination with the slit width gives a resolution of roughly 
11.8 \AA, and we show the combined spectra obtained in this 
mode in Figure~\ref{fig4:a370-6-spec}.
The strong emission line is at a wavelength of 9187 \AA,
which corresponds to a redshift of $z=6.56$ for Lyman alpha.
The continuum break is consistent with that obtained from
the imaging data but is more poorly determined because of
the uncertainties in the sky subtraction in the spectra.
\begin{inlinefigure}
\vspace{-0.07cm}
\centerline{\epsfig{file=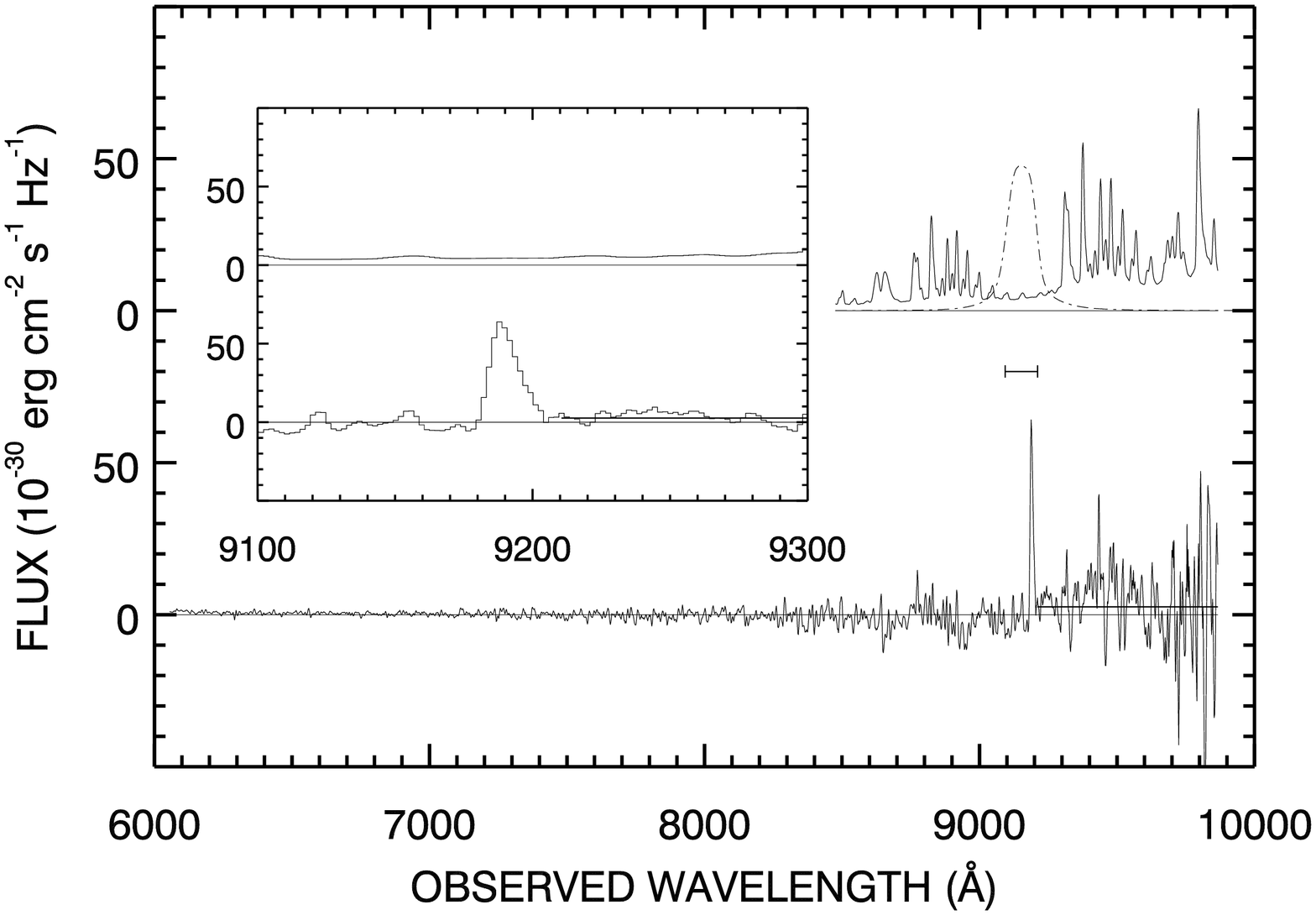,height=3.5in,angle=90}}
\caption {The spectrum of HCM 6A from a 4-hr exposure made
  with the LRIS spectrograph using the 400 line/mm grating. The
  emission line is at 9187\AA. The solid line near the axis 
  shows the median continuum flux above the line, and gives a break 
  strength consistent
  with the broadband measurements. The solid line above the emission
  shows the 50\% width of the narrowband filter, whose
  profile is shown overlaid on the fluxed nightsky spectrum, plotted
  at 1\% strength above the object spectrum. Enlarged plots 
  of the object's emission feature and 1\% of the nightsky background spectra 
  taken with the 400 line/mm grating are shown in the inset, where 
  the asymmetry of the line profile can be clearly seen.
  \label{fig4:a370-6-spec}
  }
\end{inlinefigure}

\section{Results and Discussion}

The absence of detections outside the cluster lensing regions places a 1
sigma upper limit of 27 deg$^{-2}$ for the number of sources in the filter
bandpass ($\Delta z=0.1$) at the 5 $\sigma$ flux limit of $1.6\times
10^{-17}$ erg cm$^{-2}$ s$^{-1}$ (the HDF limit) and 13 deg$^{-2}$
at $2.7\times 10^{-17}$ erg cm$^{-2}$ s$^{-1}$ (the worst limit in the
six fields), which is broadly consistent with that expected from models
\citep[e.g.,][]{thommes99} or extrapolation from lower redshifts.

We have used the Lenstool model of A370 described in \citet{kneib93}
to measure an amplification of 4.5 at the position of HCM 6A,
giving a demagnified flux of $6\times 10^{-18}$ erg cm$^{-2}$
s$^{-1}$. Lenstool also allows us to measure the corresponding source
plane areas which would give such amplifications. For the combined 
{A370}, \objectname{A851}, and \objectname{A2390} areas 
the observed source plane area is 0.46 arcmin$^2$, with most of this 
area lying behind the most massive ({A370}) cluster.
The surface density of objects above a flux limit of $4\times 10^{-18}$
erg cm$^{-2}$ s$^{-1}$ is then between 1300 and 26000 deg$^{-1}$,
where the range is $\pm1\ \sigma$, which is also broadly consistent with
the model expectations.

Equivalent width determination for very high-$z$ objects is complicated
by the strong forest absorption on the blue side of the emission feature,
and across the $Z\/$ band. For the 1 arcsec diameter aperture measurements
shown in Figure~\ref{fig3:a370-6-sed} the equivalent width determined from
the line flux and the average long-wavelength continuum flux above the 
emission feature is 190 \AA. If instead the average $Z$-band continuum is 
used as the reference value, the computed equivalent width increases to 325 \AA.

The brighter region of the object is clearly extended along an axis of
about 110$^\circ$ in PA with a maximum length of around $4''$. The
elongation is consistent with the substantial distortion expected at
this position if the galaxy lies at very high redshift, providing an
additional weak confirmation of the redshift identification. The intrinsic
size of the main region at the source plane is then around $1''$. The
second fragment lies around $1''$ away from the brighter portion.
At $z=6.56$ with the currently favored Lambda cosmology ($\Lambda=0.66$
and $\Omega=0.33$) and $H_{0}=65$ km s$^{-1}$ Mpc$^{-1}$ these angular
sizes correspond to physical sizes of approximately 40 kpc.

After removing the lensing amplification the continuum flux corresponds
to a rest frame ultraviolet luminosity of $2\times 10^{29}$ erg s$^{-1}$
Hz$^{-1}$, which may be used to estimate the star formation rate. For a
Salpeter IMF extending to 0.1 $M_{\odot}$, this is about 40 $M_{\odot}$
yr$^{-1}$. The rest frame demagnified Ly$\alpha$ luminosity is 
$2\times 10^{42}$ erg s$^{-1}$ which, assuming case B recombination
and using Kennicutt's \citeyearpar{kenn83} translation of H$\alpha$
luminosity to star formation rate, would correspond to a rate of 2
$M_{\odot}$ yr$^{-1}$. The lower value obtained from the Ly$\alpha$
luminosity presumably reflects the combined effects of dust extinction
in the galaxy and destruction of the blue side of the emission line by
scattering in the intergalactic medium.

Using the more robust UV continuum estimate we find that the universal
star formation rate at this redshift from this class of object lies in the
range $0.13\to2.7\ M_{\odot}$ Mpc$^{-3}$ yr$^{-1}$ ($\pm1\ \sigma$). This
is quite similar to rates estimated at lower redshifts within the very
large uncertainty \citep[e.g.,][and references therein]{bar00}.

Perhaps of most interest, given this is a single object and statistical
inferences are therefore difficult, is that the presence of even a single
object at this redshift may suggest we have not yet reached the redshift
of reionization. In a pre-reionization epoch the radiation damping
wings from the neutral gas will black out regions extending about 20
\AA\ to the redward of the Ly$\alpha$ wavelength in the observed frame
of the object \citep{jordi,jordi2}.
While ultraluminous quasars can ionize their surroundings and prevent
this effect, it is unlikely that a small Ly$\alpha$ emitting galaxy of
the present type could do this though it could
lie in a large ionized region produced by some other source of
ionizing photons.

\acknowledgements
This work was supported in part by the State of Hawaii and by NSF grant
AST-0071208 and NASA grant GO-7266.01-96A from Space Telescope Science 
Institute, which is operated by AURA, Inc., under NASA contract NAS 5-26555.  
RGM thanks the Raymond and Beverly Sackler Foundation for support.  We thank 
Amy Barger for taking additional LRIS $R$-band images of A370 to improve our
non-detection limits.

\clearpage

\end{document}